\documentclass{mem}
\usepackage{natbib}
\usepackage{txfonts}
\usepackage{balance}
\usepackage{graphicx}
\usepackage{hyperref}
\usepackage{url}
\bibpunct{(}{)}{;}{a}{}{,} 
\usepackage{scrextend}
\hypersetup{
		colorlinks, 
		citecolor=blue, 
		filecolor=blue, 
		linkcolor=blue,
		breaklinks=true, 
		plainpages=false,
		urlcolor=blue 
}
\title{Enhanced methods for computing spectra from CO5BOLD models using Linfor3D}
\subtitle{Molecular bands in metal-poor stars}

\author{A. J. Gallagher\inst{1}\thanks{Observatoire de Paris fellow}
\and
M. Steffen\inst{2,1}
\and
E. Caffau\inst{1}
\and
P. Bonifacio\inst{1}
\and 
H.-G. Ludwig\inst{3,1}
\and
B. Freytag\inst{4}
}
\institute{GEPI, Observatoire de Paris, PSL Research University, CNRS, Univ. Paris Diderot, 
Sorbonne Paris Cit\'{e} Place Jules Janssen, 92190 Meudon, France.\\ 
email: \texttt{andrew.gallagher@obspm.fr}
\and 
Leibniz-Institut f{\"u}r Astrophysik Potsdam, An der Sternwarte 16, 14482 Potsdam, Germany.
\and 
Zentrum f{\"u}r Astrononmie der Universit{\"a}t Heidelberg, Landessternwarte, K{\"o}nigstuhl 12, 69117 Heidelberg, Germany.
\and
Department of Physics and Astronomy at Uppsala University, Regementsv{"a}gen 1, Box 516, SE-75120 Uppsala, Sweden.
}
\authorrunning{A. J. Gallagher et al.}
\titlerunning{Enhanced Linfor3D computations}
\newcommand{\cobold}{CO$^{5}$BOLD}

\newcommand{\odx}{{\tt LHD}}
\newcommand{\teff}{T_{\rm eff}}
\newcommand{\logg}{\log{g}}

\newcommand{\feh}{{\rm [Fe/H]}}

\newcommand{\ca}{A({\rm C})}

\newcommand{\na}{A({\rm N})}

\newcommand{\oa}{A({\rm O})}

\newcommand{\ppaper}{Paper I}
\abstract{
Molecular features such as the G-band, CN-band and NH-band are important diagnostics for measuring a star's carbon and nitrogen abundances, especially in metal-poor stars where atomic lines are no longer visible in stellar spectra. Unlike atomic transitions, molecular features tend to form in bands, which cover large wavelength regions in a spectrum. While it is a trivial matter to compute carbon and nitrogen molecular bands under the assumption of 1D, it is extremely time consuming in 3D. In this contribution to the 2016 \cobold\ workshop we review the improvements made to the 3D spectral synthesis code Linfor3D, and discuss the new challenges found when computing molecular features in 3D.
}
\begin{document}
\maketitle
\section{Introduction}
\label{sec:introduction}
When the metallicity of a star falls below [Fe/H]=-2.0, the only probes of the star's carbon abundance (in the visible part of the spectrum) are found in molecular features, which form in large bands in the stellar spectrum. While computation of carbon and nitrogen molecular bands for the purposes of modelling is a relatively simple matter in 1D, it is extremely time-consuming in 3D. This presents difficulties when analysing a well-known sub-class of metal-poor stars known as carbon-enhanced metal-poor (CEMP) stars, as the carbon abundances are determined through exploiting molecular carbon features, like the G-band.

During the 2012 \cobold\ workshop, it was highlighted that one of the main spectrum synthesis codes used with \cobold, Linfor3D \citep{Steffen2015}\footnote{\href{http://www.aip.de/Members/msteffen/linfor3d}{http://www.aip.de/Members/msteffen/linfor3d}}, would benefit from a sizeable upgrade so that large spectral ranges, such as molecular band features, could be computed in a reasonable amount of time \citep[][henceforth \ppaper]{Bonifacio2013}.

This contribution to the 2016 \cobold\ workshop highlights the major improvements made to Linfor3D since the 2012 \cobold\ workshop that address the issues raised in \ppaper.

\section{Computing large wavelength regions with Linfor3D}
\label{sec:linfor}

Over the past three years, we have made significant efforts to improve Linfor3D with a primary focus of optimising it to handle large wavelength regions, and compute them in a reasonable amount of time. This has allowed us to compute large molecular bands, like the G-band, for the first time with Linfor3D \citep{Gallagher2016}.

\subsection{What is new?}
\label{sec:nlinfor}

The latest versions of Linfor3D (versions 6.0.0 onwards) have been ported to GNU Data Language\footnote{\href{http://gnudatalanguage.cvs.sourceforge.net/}{http://gnudatalanguage.cvs.sourceforge.net/}\label{gdldb}} (GDL) from Interactive Data Language (IDL), but it still maintains the ability to run in IDL. In fact, many of the advanced plotting options available during a Linfor3D simulation \citep{Steffen2015} are currently only executable when running with IDL. This is because GDL is still very much under development, and the plotting packages, in particular those relating to post scripts, are either unavailable, or not advanced enough. We stress that only the \emph{CVS version} of GDL\footref{gdldb} is currently compatible with Linfor3D, while the version downloadable through the RPM Package Manager is missing crucial libraries necessary to successfully run Linfor3D. New routines that handle the Input/Output (I/O) were written for Linfor3D. They take advantage of the Universal Input Output (UIO) library written by B. Freytag\footnote{\href{http://www.astro.uu.se/~bf/co5bold_main.html}{http://www.astro.uu.se/$\sim$bf/co5bold\_main.html}}, but were designed to mimic the call sequences used by the internal IDL SAVE/RESTORE routines. As such, all I/O data in Linfor3D are now in UIO format (idlsave $\rightarrow$ uiosave).

Computation of any 3D spectrum is time consuming. The aim of this upgrade was to expand the capabilities of Linfor3D so that it can produce large amounts of 3D spectra quickly. With its new ability to run on GDL, where expensive license concerns are moot, Linfor3D can be run on High Performance Computing (HPC) centres and can utilise parallel processing. Although this can only currently be done as an ``embarrassingly parallel'' job, owed to the limitations of IDL and GDL in multi-core processing. To that end, we have developed a pipeline around Linfor3D for use on HPC centres. It is designed to handle the task of setting up Linfor3D simulations, submitting them together in parallel (under MPI), monitor them and combine them upon completion. For the moment, this pipeline is not available as it is still under development. However, it will be released with future versions of Linfor3D once development concludes.

\begin{figure*}[!t]
\begin{center}
\includegraphics[width=\linewidth]{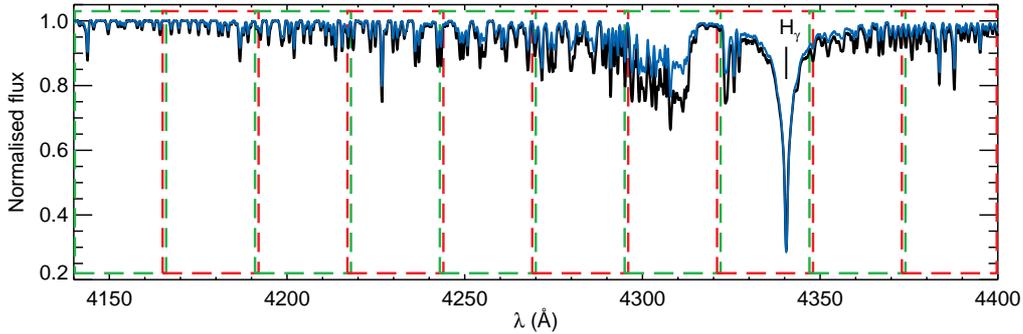}
\caption{The G-band in 3D (black) and 1D (blue) for a typical dwarf star with \cobold\ name designation {\tt d3t63g40mm30n02}. Linfor3D jobs were run in parallel over wavelength intervals (red and green boxes) and snapshots ($10\times20$ jobs).}
\label{fig:gband}
\end{center}
\end{figure*}

The pipeline takes a large wavelength range, like the G-band, and splits it into several wavelength intervals. Each wavelength interval is then computed for every \cobold\ model snapshot selected. At this stage, the user has the choice of binning the separate Linfor3D jobs together and computing them as a single multi-core job using MPI. There is no limit on the number of jobs that are grouped together, providing that the correct number of compute nodes and CPUs are defined, as the jobs remain independent of one-another. We have successfully tested this for upto 48 separate Linfor3D jobs on four 12-core compute nodes on an HPC centre. This means that one can submit an entire band, consisting of $N$ Linfor3D jobs as a single parallel session of $N$ CPUs (if compute resources are available), making it far quicker to compute a large wavelength range. In fact, Linfor3D will now compute the G-band molecular feature ($\sim4308$ atomic and molecular transitions and ${\rm H}_\gamma$) for a 20 snapshot 3D model atmosphere in 24 hours (on 200 CPUs running in parallel), whereas it is estimated that this would take over 90 days of compute time when run sequentially.

Upon completion, the pipeline runs a series of checks on the output data. These checks test  whether the output data exists, whether the data can be opened and whether the integrity of the output data was maintained. If this check does not return an error, the data are combined into a single, standard Linfor3D output, formatted as a UIO, which can be used in IDL and GDL. If, however, there are problems with the output data (missing or corrupt files), the pipeline determines which data are missing, and resubmits them to the HPC centre.

Finally, Linfor3D now has a new run mode that allows it to be executed as a stand-alone 1D spectrum synthesis code. This is particularly useful if large grids of 1D syntheses are needed to fit the 3D when the 3D synthetic spectra have been broken up into smaller wavelength intervals, as was done for the G-band (Sect.~\ref{sec:gband}). This prevents the introduction of new systematic errors created when comparing spectra computed using different spectrum synthesis codes. Linfor3D still retains the ability to compute spectra based on 1D MARCS \citep{Gustafsson2008}, ATLAS \citep{Kurucz2005}, and \odx\ \citep{Caffau2007} model atmospheres. Further details of this new feature are presented in the Linfor3D user manual, \citet{Steffen2015}.

\section{Computing molecular bands}

Using the pipeline, we have been able to construct a grid of Linfor3D data for the G-band and CN-band under the assumption of local thermodynamic equilibrium (LTE). We now describe the details in the construction of this grid and detail the methods used to compute the bands.

\subsection{The G-band}
\label{sec:gband}

The  $A^{2}\Delta - X^{2}\Pi$ CH molecular feature, commonly referred to as the G-band consists of $3353$ CH molecular and, for our work, upto 955 atomic transitions, as well as the ${\rm H}_\gamma$ Balmer feature, over a $260$\,\AA\ wavelength range between $4140-4400$\,\AA. Further details of the line lists used for our work are given in \citet{Gallagher2016}. As such, computing this feature sequentially in Linfor3D is not practical, as was foreshadowed in \ppaper. Using the pipeline just described, we are able to fully compute the G-band in sufficient detail. We have constructed a sizeable grid of G-band spectra with Linfor3D, covering the majority of the \cobold\ grid. At the current date we have $500$ synthetic G-band spectra for $70$ \cobold\ model atmospheres over the metallicity range $-4.0\leq\feh\leq-0.5$.

Every \cobold\ model used to construct the G-band grid has at least seven CNO abundances. Six of these CNO abundances are computed for a solar CNO mixture, which is scaled by the carbon abundance. The synthetic G-band of every model atmosphere was computed over a carbon abundance range $6.0\leq\ca\leq8.0$, \emph{except} for those models with $\feh=-2.0$, where we compute spectra over the carbon abundance range $6.5\leq\ca\leq8.5$. All carbon abundances were computed over an interval $\Delta\ca=0.4$\,dex. We also compute a single synthetic spectrum with a reduced oxygen abundance, and hence an enhanced C/O\footnote{${\rm X/Y}=N({\rm X})/N({\rm Y})=10^{[A({\rm X})-A({\rm Y})]}$} ratio, for every model atmosphere used \citep[precise values for these syntheses given in][]{Gallagher2016}. This was done to address the effect the C/O ratio has on CH transitions; the higher the C/O ratio, the stronger the CH transitions become. Full details are available in \citet{Gallagher2016}. The G-band grid is expanding. We intend to reduce the abundance interval between the CNO abundances and expand the range in $\ca$. This will provide a very large grid of syntheses to utilise in CEMP star analyses, which will be made public.

\begin{figure}[!t]
\begin{center}
\includegraphics[width=\linewidth]{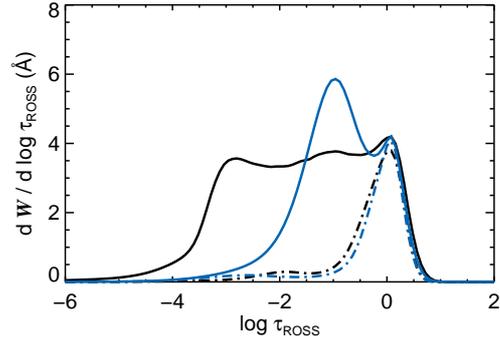}
\caption{The contribution functions of the G-band in 3D (black) and 1D (blue) for the data depicted in Fig~\ref{fig:gband}. The 3D (black) and 1D (blue) ${\rm H}_\gamma$ contribution functions have also been included (dashed-dot lines).}
\label{fig:cfgband}
\end{center}
\end{figure}

Figure~\ref{fig:gband} presents the G-band for a typical dwarf turn-off star with stellar parameters $\teff/\logg/\feh=6240/4.0/-3.0$ ({\tt d3t63g40mm30n02}). This model has been synthesised for $\Delta A({\rm CNO})=+2$\,dex, such that $\ca=7.39$, $\na=6.78$ and $\oa=7.66$. The green and red boxes graphically represent the wavelength intervals used during the synthesis. Each interval was 26\,\AA\ wide plus a 1\,\AA\ synthesis overlap with its neighbour. The continuum of each interval was computed using three points inside each interval, which were the minimum and maximum wavelengths of the interval and the central wavelength of the interval. Substantial testing showed that the overlapping regions of the G-band synthesis were perfectly reproduced by the intersecting wavelength interval. The counterpart 1D grid used for analysis in \citet{Gallagher2016} was constructed using the new run mode in Linfor3D (Sect.~\ref{sec:nlinfor}), but it was not synthesised using the wavelength intervals used during the 3D synthesis. Instead, each 1D synthetic profile was computed for the entire G-band (260\,\AA). Test comparisons between them and 1D syntheses that were computed over the wavelength intervals demonstrated no discernible differences. This is because the continuum over the G-band wavelength range is fairly monotonic, which means that the three points used to compute the continuum in Linfor3D (as described above) were sufficient.

\begin{figure*}[!ht]
\begin{center}
\includegraphics[width=\linewidth]{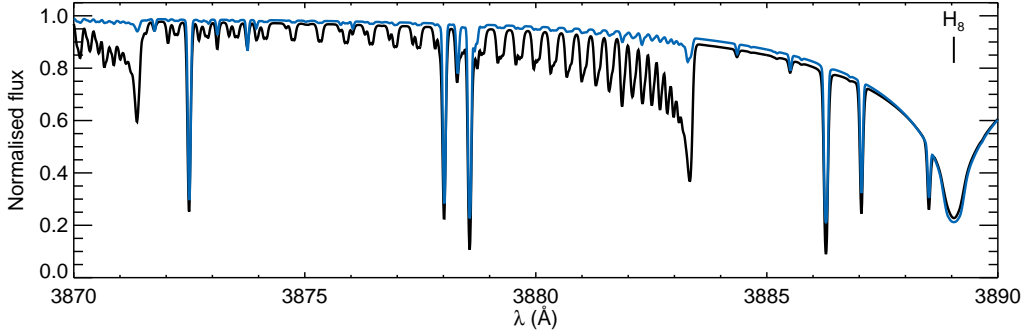}
\caption{The CN-band in 3D (black) and 1D (blue) for a typical dwarf star with \cobold\ name designation {\tt d3t63g40mm30n02}. Linfor3D was run in parallel over the snapshots (20 snapshots) for the entire wavelength range.}
\label{fig:cnband}
\end{center}
\end{figure*}

It is evident on inspection that the lines computed by the 3D synthesis are stronger than those computed in the counterpart 1D (which was computed using the \odx\ model for this \cobold\ model). Full details of the reasons behind this are found in \citet{Gallagher2016}. In short, the cooler temperatures in the 3D model leads to enhanced molecule formation in the outer regions of the model, under LTE. This is presented graphically in Fig.~\ref{fig:cfgband} as a contribution function. As depicted, the 3D and its equivalent 1D syntheses form in different regions of their respective atmospheres. The integral of the two contribution functions also provide the total equivalent width of the feature. By eye, it is evident that the strength of the 3D G-band is larger than its 1D counterpart. The peak that the 1D and 3D syntheses share in the deeper regions of the models are primarily due to the core of the ${\rm H}_\gamma$ Balmer line, which is shown by the dashed-dot plots in the figure.

\begin{figure}[!t]
\begin{center}
\includegraphics[width=\linewidth]{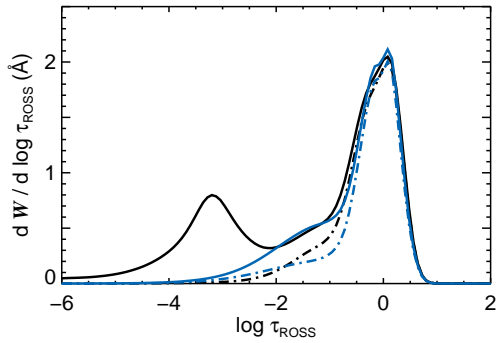}
\caption{The CN-band in 3D (black) and 1D (blue) for the data depicted in Fig~\ref{fig:cnband}. The 3D (black) and 1D (blue) ${\rm H}_8$ contribution functions have also been included (dashed-dot lines).}
\label{fig:cfcnband}
\end{center}
\end{figure}

\subsection{The CN-band}
\label{sec:cnband}

\ppaper\ examined the $B^{2}\Sigma^{+}−X^{2}\Sigma^{+} (0-0)$ bimetallic molecular feature commonly referred to as the CN-band. It was stated that this feature, in comparison to larger molecular features, is much easier to compute. Using the pipeline described above, we are able to compute the CN-band, which covers the wavelength range $3870-3890$\,\AA, in approximately 24 hours. The molecular data for this band was taken from \citet{Masseron2014}\footnote{The database is available at \href{http://www.as.utexas.edu/~chris/lab.html}{http://www.as.utexas.edu/$\sim$chris/lab.html}.} and the atomic data was taken from the ``Turn-Off Primordial Stars'' (TOPoS) ESO/VLT large programme 189.D-0165 \citep{Caffau2013}, which was in turn taken from the ``First Stars'' ESO large programme 165.N-0276(A) \citep{Cayrel2004}. The list includes $871$ CN molecules, $47$ atomic transitions and the ${\rm H}_8$ and ${\rm H}_9$ Balmer lines. Only the ${\rm H}_8$ feature is visible in the selected wavelength range. The ${\rm H}_9$ Balmer line forms at $3835$\,\AA\ but the wings of the line extend much further. An example of the CN-band is presented in Fig.~\ref{fig:cnband}, which was computed using the same CNO abundances as was used when computing the G-band in Fig.~\ref{fig:gband}. Like the G-band, the CN-band presents large differences between the overall strength of the 3D and 1D profiles. That will lead to large 3D corrections, reducing the carbon and nitrogen abundance in a star. Fig.~\ref{fig:cfcnband} illustrates that the CN-band feature also forms in the outer regions of 3D model atmosphere. This is due to temperature and density fluctuations of the 3D model, which a 1D model cannot replicate, as well as the cooler overall temperature of the 3D model, relative to the counterpart 1D model. In this example, the 1D and 3D formation is dominated by the ${\rm H}_8$ Balmer feature (represented by the dashed-dot lines) in the deeper regions of the atmospheres.

\begin{figure*}[!ht]
\begin{center}
\includegraphics[width=\linewidth]{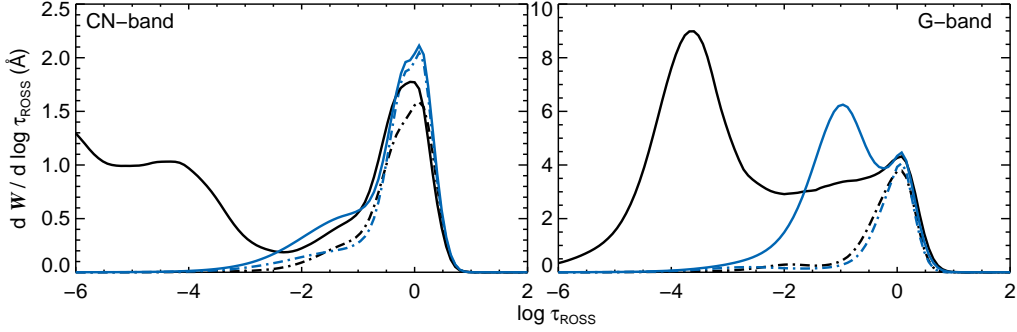}
\caption{The effect of the increasing C/O ratio on the CN-band (left) and G-band (right) molecular features in 3D (black) and 1D (blue). ${\rm C/O}=21.4$ in both bands.}
\label{fig:hico}
\end{center}
\end{figure*}

As the wavelength range is considerably smaller than the G-band, it was unnecessary to compute the feature in wavelength intervals. However, we do compute synthesis for the snapshots in parallel (typically 20 syntheses in parallel). This makes the CN-band an ideal feature to analyse the C or N abundance in a star (in conjunction with the G- or NH-band). However, there is a problem.

\subsubsection{The C/O ratio}
\label{sec:coratio}

The formation of the CN-band in a stellar atmosphere is complex. In fact, it has been known for 40 years that the solar CN-band is extremely sensitive to departures from thermodynamic equilibrium (NLTE), and that part of it forms in the chromosphere \citep{Mount1975}. It has also been confirmed that metal-poor dwarf stars have chromospheres \citep{Takeda2011,Smith2012}, suggesting that the same stipulations apply to CN-band formation in metal-poor regimes. As such, it would be important to model the chromosphere when computing the CN-band, even in metal-poor dwarf stars, which includes the majority of known CEMP stars. Linfor3D is not currently capable of computing synthesis with \cobold\ models that include a chromosphere \citep{Wedemeyer2013}, until another substantial upgrade is applied. 

\ppaper\ demonstrated that the C/O ratio has a severe effect on the formation of the CN-band in 3D; a larger C/O ratio pushes formation to the outer regions of the computational box, and when the C/O ratio exceeds $1.0$, a significant fraction of the band forms outside the confines of the 3D computational box.

Figures~\ref{fig:cfgband}~and~\ref{fig:cfcnband} do not demonstrate any unusual behaviours in the CN- or G-band. The bands are forming inside the confines of the computational box, and the contributions at different optical depths can be explained by the different components that make these features up, i.e. atomic and molecular transitions and the Balmer series. However, the C/O ratio is small in these examples, ${\rm C/O}=0.54$. As the C/O ratio is increased to $21.4$ (Fig.~\ref{fig:hico}), the problems presented in \ppaper\ can be reproduced. In this example, $A({\rm CN})$ have been enhanced, but oxygen has not, such that $\ca=7.39$, $\na=6.78$ and $\oa=6.06$. The effect on formation in both bands is severe, when compared to those depicted in Figs.~\ref{fig:cfgband}~\&~\ref{fig:cfcnband}, but the problems are worse in the CN-band, as the majority of its formation is pushed outside the confines of the computational box. A complete description of why the C/O ratio has this effect on the carbon bearing molecules is given in \citet{Gallagher2016}

We are currently investigating scenarios to correct for the 3D-LTE effect of CN and CH formation, however, this will be reported in a future paper. Nevertheless, it is clear that the C/O ratio is an important parameter for the CN- and G-band, and hence a reasonable estimate of the oxygen abundance would be ideal to resolve the problem in every star. However, this is not possible for most metal-poor stars, so another solution should be pursued.

\begin{figure*}[!t]
\begin{center}
\includegraphics[width=0.85\linewidth]{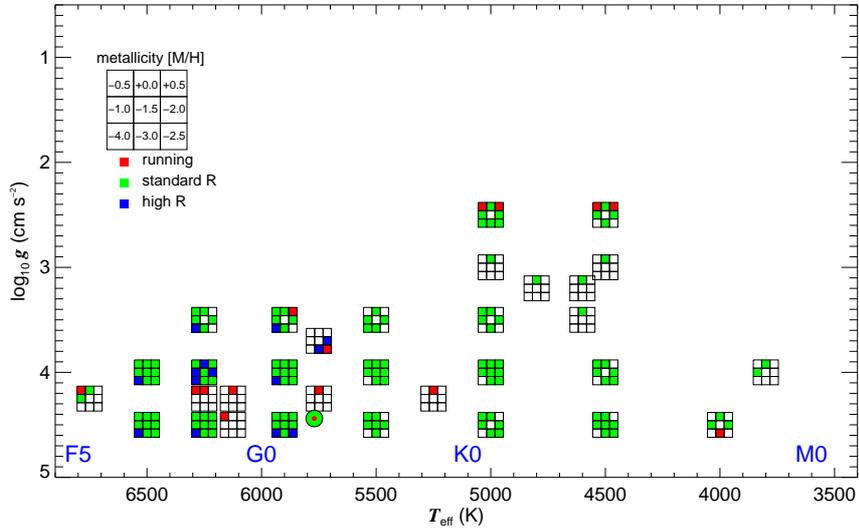}
\caption{The current state of the CIFIST \cobold\ grid as of the date of this publication. We only include models for which we are interested in computing molecular bands.}
\label{fig:hydrogrid}
\end{center}
\end{figure*}

\subsection{The NH-band}
\label{sec:NHband}

The NH $A^{3}\Pi_{i}-X^{3}\Sigma^{-}$ band (3360\,\AA\ NH-band) was also studied in \ppaper. However, the size of the line list was severely reduced so that its computation with the older version of Linfor3D could be completed within a reasonable time frame. We have not attempted to compute the NH-band for this work. Instead, we provide estimates based on the work presented in \ppaper\ and work done on other bands, presented here.

The original NH-band line list utilised in \ppaper, from the Kurucz list\footnote{\href{http://kurucz.harvard.edu/LINELISTS/LINESMOL/nh.asc}{http://kurucz.harvard.edu/LINELISTS/\newline LINESMOL/nh.asc}}, consisted of 614 NH transitions over an $11$\,\AA\ range ($3356-3366$\,\AA). The line list was down-selected so that only 138 NH lines were considered, as well as nine atomic lines, and the band was computed over two wavelength intervals. With the upgrades made to Linfor3D, \emph{the original} line list (614 NH transitions) could be computed in 24 hours, when the syntheses are run in parallel over the snapshots. However, this does not account for the entire NH-band around $3345-3375$\,\AA. There are $1375$ NH molecular transitions and $\sim212$ atomic transitions in the Kurucz database, totalling $1577$ lines over $30$\,\AA. Based on the work done in \citet{Gallagher2016}, we estimate that such a band would require slightly more than 24 hours to compute, if the computations were run in parallel over the snapshots only. To reduce this time, it would be useful to split the band into two wavelength intervals (similar to what was done with the G-band) and recombine them upon completion. As the pipeline does this automatically, such an undertaking would not be difficult.

It has been shown in \ppaper\ and \citet{Gallagher2016} that a line list can be down-selected to suit the star's metallicity, which would also speed up the synthesis considerably, since Linfor3D requires the same time to compute a weak or strong line. Therefore, this list could be reduced, thereby shortening the time taken for its computation. Nevertheless, it appears that such work is feasible with the latest versions of Linfor3D and grids will be computed in the future.

\section{The \cobold\ model grid}
\label{sec:coboldgrid}

The \cobold\ community is very active in the generation of new models that span the entire H-R diagram. The Cosmological Impact of the FIrst STars \citep[CIFIST -- ][]{Ludwig2009} collaboration has been computing models that cover the stellar parameters depicted in Fig.~\ref{fig:hydrogrid}. The \cobold\ community is heavily invested in the computation of \cobold\ models for giants, super giants, white dwarfs, and brown dwarfs, as well as \cobold\ models that include a chromosphere. As such, the H-R diagram provided here is vastly under-represented. A number of the \cobold\ models in Fig.~\ref{fig:hydrogrid} have been computed several times, for various reasons (new opacities, different opacity binnings, different scattering treatments, etc.). Nevertheless, the green marks depict the models that have been computed for a geometrical resolution of $140\times140\times150$ grid points, and those marked in blue have been computed for a higher geometrical resolution of $280\times280\times300$ grid points. Models computed for the smallest geometrical resolution have been computed for 5, 6, 12 or 14 bins -- most have 12 or 14 opacity bins, however. All models with the higher geometrical resolution have been computed with 12 to 14 opacity bins. A detailed description of the opacity binning is provided in \citet{Ludwig2013}. Boxes marked in red represent models that are currently being computed, while those in white have yet to be computed. We stress that these models represent a part of the combined \cobold\ grid. 

\begin{acknowledgements}
This project is funded by FONDATION MERAC and the matching fund granted by the Scientific Council of Observatoire de Paris. We acknowledge support from the Programme National de Cosmologie et Galaxies (PNCG) and Programme  National de Physique Stellaire (PNPS) of the Institut National de Sciences de l'Univers of CNRS.
This work was granted access to the HPC resources of MesoPSL financed by the Region Ile de France and the project Equip@Meso (reference ANR-10-EQPX-29-01) of the programme Investissements d’Avenir supervised by the Agence Nationale pour la Recherche.
This work was supported by Sonderforschungsbereich SFB 881 ``The Milky Way System'' (subproject A4) of the German Research Foundation (DFG).
\end{acknowledgements}

\bibliographystyle{aa}

\end{document}